\documentclass[preprint,prc,showpacs,preprintnumbers,amsmath,amssymb,floatfix]{revtex4}
\usepackage{amsmath}
\usepackage{graphicx}
\usepackage{dcolumn}
\usepackage{bm}
\usepackage{ulem} 
\usepackage[usenames]{color}
\usepackage{epstopdf}
\usepackage{epsfig}
\usepackage{float}
\usepackage{subfigure}
\usepackage{multirow}
\usepackage[version=3]{mhchem}

\newcommand{\nc}{\newcommand}       
\nc{\vc}[1] {\mbox{\boldmath $#1$}} 
\nc{\del}       {\partial}              
\nc{\bra}       {\langle}               
\nc{\ket}       {\rangle}               
\nc{\bras}[1]   {\langle #1|}           
\nc{\kets}[1]   {|#1\rangle}            
\nc{\mapleft}[1]{           
 \smash{\mathop{\,          %
  \hbox to 1.5cm{\rightarrowfill}\, }\limits_{#1}}}
\nc{\beq}     {\begin{eqnarray}} \nc{\eeq}    {\end{eqnarray}}
\nc{\nn}      {\\\nonumber} \nc{\vs}      {\vspace{-0.275cm}}
\nc{\fra}    {\frac{1}{2}}
\nc{\mb}        {\mathbf}

\begin{document}


\title{Massive neutron star with strangeness in a relativistic mean field model with a high-density cut-off}

\author{Ying Zhang}
\affiliation{Department of Physics, Faculty of Science, Tianjin University, Tianjin 300072, China}

\author{Jinniu Hu}
\email{hujinniu@nankai.edu.cn}
\affiliation{School of Physics, Nankai University, Tianjin 300071, China}

\author{Peng Liu}
\affiliation{Tianjin Institute of Aerospace Mechanical and Electrical equipment, Tianjin 300301, China}

\date{\today}
\begin{abstract}
The properties of strangeness neutron star are studied within relativistic mean field (RMF) model via including a logarithmic interaction as a function of scalar meson field. This logarithmic interaction, named as the $\sigma$-cut potential, can largely reduce the attractive contributions of scalar meson field at high density without any influence on nuclear structure around normal saturation density. In this work, the TM1 parameter set is chosen as the RMF interaction, while the strengths of logarithmic interaction are  constrained by the properties of finite nuclei so that we can obtain a reasonable effective nucleon-nucleon interaction. The hyperons, $\Lambda,~\Sigma$, and $\Xi$ are also considered in neutron stars within this framework, whose coupling constants with mesons are determined by the latest hyperon-nucleon and $\Lambda$-$\Lambda$ potentials extracted from the experimental data of hypernuclei. The maximum mass of neutron star can be larger than two solar mass with  these hyperons. Furthermore, the nucleon mass at high density will be saturated due to this additional $\sigma$-cut potential, which is consistent with the conclusions from the microscopic calculations such as, Brueckner-Hartree-Fock theory and quark mean field model. 
\end{abstract}



\maketitle

\section{Introduction}
The research objects in nuclear physics are very compact condensed matter, where the nucleons interact with each other through the nuclear force as an effective interaction of QCD theory at low energy scale~\cite{oertel17}. Due to the complication of nuclear force, there is still not a uniform theory, which can describe the properties of all nuclei in the nuclide chart perfectly. However, with the development of computer technology and nuclear many-body methods, the properties of finite nuclei in the nuclide chart can be simulated reasonably by various {\it ab initio} calculation methods~\cite{carlson15,dickhoff04,hagen14a,lee09,meissner15,barrett13,shen16}, shell model~\cite{caurier05}, and density functional theory (DFT)~\cite{bender03,stone07,ring96,vretenar05,meng06,niksic11} at different mass regions from light to heavy nuclei.  

The DFT in nuclear physics is constructed based on an effective nucleon-nucleon ($NN$) interaction, which is determined by fitting the empirical saturation properties of infinite nuclear matter, or the ground-state properties of several stable nuclei.  The $NN$ interaction is expressed as a function of nuclear density in DFT. The earliest available DFT in nuclear physics, Skyrme-Hartree-Fock (SHF) model, was developed in 1970s with a non-relativistic zero-range $NN$ interaction proposed by Skyrme based on the mean-field approximation~\cite{vautherin72}. Later, the covariant version of DFT was realized in nuclear physics by Walecka through introducing the exchanges of scalar and vector mesons between different nucleons~\cite{walecka74}. A lot of advanced DFT have been proposed until now, which can successfully describe the properties of majority nuclei discovered in experiments~\cite{bender03,stone07,ring96,vretenar05,meng06,niksic11} . 

When these DFT nuclear forces were applied to investigate the properties of infinite nuclear matter, identical behaviors of their equations of state (EOSs) were obtained at low density (around the nuclear saturation density, $\rho_0\sim0.15$ fm$^{-3}$), while the EOSs from DFT at high density were quite different~\cite{dutra12,dutra14}. It is easily understood that the strengths of DFT nucleon forces are strongly related to the experimental data of nuclear many-body system closed to saturation density region, but the constraint of experimental information at high density is rather few. In present status, the compact matter was only generated up to $2\rho_0\sim3\rho_0$ mostly in laboratory from heavy ion collision~\cite{danielewicz03}.   However, a lot of investigations made out that it approaches $5\rho_0\sim10\rho_0$ in the core region of compact star in the universe~\cite{oertel17}.  Therefore, the DFT has ambiguity when it is applied on the study of neutron stars.

Actually, {\it ab initio} calculation methods are  good candidates to work out the properties of nuclear matter at high density, which adopt the realistic $NN$ interactions from the $NN$ scattering data. Without the three-body nucleon force, the available non-relativistic  {\it ab initio} methods cannot reproduce the empirical saturation properties completely~\cite{li06,hu16}. Once the three-body effect was included~\cite{li08}, the high density behaviors were in accordance with those from the relativistic  {\it ab} methods, such as relativistic Brueckner-Hartree-Fock (RBHF) theory~\cite{brockmann90}.

With the strangeness degree of freedom, the shortcoming of DFT becomes obvious, especially in neutron star. A lot of DFT interactions with hyperons produce too soft EOSs to obtain  massive neutron stars, which were confirmed recently from astronomical observables, i.e. PSR J1614-2230  ($1.928\pm0.017 M_\odot$)~\cite{demorest10,fonseca16} and PSR J0348+0432 ($2.01\pm0.04 M_\odot$)~\cite{antoniadis13}. Therefore many mechanisms were introduced to make the EOS become stiffer at high density region, e.g. the repulsive components of hyperon-hyperon force~\cite{weissenborn12,oertel15}, three-body hyperon-nucleon force~\cite{yamamoto13}, and quark phase~\cite{demorest10}. However, these considerations will more or less influence the properties of nuclear matter around the saturation density at the same time.

A few years ago, Maslov {\it et al.} proposed a "$\sigma$-cut" term in the Lagrangian of relativistic mean field (RMF)
model~\cite{maslov15,kolomeitsev17}, one version of covariant DFT theory,  to prevent the scalar field decreasing monotonically with the nucleon density  so that the EOS of nuclear matter at high density become stiff enough to generate massive neutron stars.  This $\sigma$-cut potential only plays its role when the density is larger than a certain value. Therefore, the scheme does not effect the properties of nuclear matter at low density.  It means that if the strength of  $\sigma$-cut potential is chosen properly, this new Lagrangian can not only provide the stiff EOS but also describe the properties of finite nuclei very well with the original parameters of RMF model. In the work of Maslov {\it et al.}~\cite{maslov15}, they just discussed this $\sigma$-cut scheme in nuclear matter. Later, Dutra {\it et al.} adopted this framework to make several available RMF interactions satisfy the constraint of massive neutron stars~\cite{dutra16}.  In this work, we would like to extend this method to finite nuclei system and neutron star with strangeness degree of freedom. Through the constraints of massive neutron stars and the experiment data of finite nuclei, the strength of $\sigma$-cut potential will be determined reasonably.

The paper is arranged as follows. In Sec. II, we briefly introduce the formulas of RMF model with $\sigma$-cut interaction. In Sec. III, the properties of several doubly-magic nuclei, nuclear matter, and neutron stars with hyperons will be shown numerically to discuss the strengths of $\sigma$-cut potential. Finally, a summary is given in Sec. IV.

\section{Formalism}
In the picture of RMF model, the baryons interact with each other by exchanging the mesons in different mass regions, like scalar meson ($\sigma$) and vector mesons ($\omega$ and $\rho$), therefore, the Lagrangian of RMF model can be expressed by the baryon fields, $\psi_B$ and meson fields as~\cite{maslov15,shen02}, 
\beq
{\cal L}
&=&
\sum_B\bar\psi_B\left[ i\gamma_\mu\partial^\mu-(M_B+g_{\sigma B}\sigma)
-g_{\omega B}\gamma^\mu \omega_\mu 
-g_{\rho B} \gamma^\mu\vec\tau_B\cdot\vec\rho_\mu 
-e\frac{(1+\tau_{B,3})}{2}\gamma^\mu A_\mu 
\right]\psi_B\nn
&&
+\frac{1}{2} \partial^\mu\sigma\partial_\mu\sigma
-\frac{1}{2} m_\sigma^2\sigma^2
-\frac{1}{3} g_2\sigma^3
-\frac{1}{4} g_3\sigma^4-U_{\text{cut}}(\sigma)\nn
&&
-\frac{1}{4} W^{\mu\nu}W_{\mu\nu}
+\frac{1}{2} m_\omega^2\omega^2
+\frac{1}{4} c_3\omega^4\nn
&&
-\frac{1}{4} \vec R^{\mu\nu}\vec R_{\mu\nu}
+\frac{1}{2} m_\rho^2\rho^2
-\frac{1}{4}F^{\mu\nu}F_{\mu\nu},
\eeq
where the arrows denote the isospin vectors of $\rho$ meson and three tensor operators for the vector and photon fields are defined as follows,
\beq
W^{\mu\nu}&=&\partial^\mu\omega^\nu-\partial^\nu\omega^\mu,\nn
\vec R^{\mu\nu}&=&\partial^\mu\vec\rho^\nu-\partial^\nu\vec\rho^\mu,\nn
F^{\mu\nu}&=&\partial^\mu A^\nu-\partial^\nu A^\mu.
\eeq 
The $\sigma$-cut potential is adopted as the logarithmic form following the work of Maslov {\it et al.}, which only influences  the $\sigma$ field at high density,
\beq
U_{\text{cut}}(\sigma)=\alpha\ln\{1+\exp[\beta(g_\sigma\sigma/M_N-f_{s})]\},
\eeq
where, $\alpha=m^4_\pi$ and $\beta=120$ to ensure the EOS being stiffer at high density~\cite{maslov15}. The factor $f_s$ is a free parameter in this work whose magnitude will be decided by the properties of finite nuclei and massive neutron stars. The larger $f_s$ leads to the $\sigma$-cut potential working from higher density. Here, we concentrate on the study of doubly-magic nuclei, which are treated as spherical cases and the spatial components of vector mesons will become zero due to the time-reversal symmetry. Hence, only time components of $\omega,~\rho$ and $A$ fields exist. For the convenient presentation later on, we would like to use the symbols, $\omega,~\rho,~A$,  instead of $\omega^0,~\rho^0,~A^0$.

The equations of motion about baryons and mesons can be generated from the Euler-Lagrange equations. However, in these equations of motion, the quantum fields cannot be solved exactly to many-body system. The mean-field approximation and no-sea approximation are taken into account to treat the mesons as classical fields in RMF model. Then, the Dirac equations of baryons are written as,
\beq
&&\left[i\gamma_{\mu}\partial^{\mu}-(M_B+g_{\sigma B})
-g_{\omega B}\omega\gamma^0
-g_{\rho B}\rho\tau_{B,3}\gamma^0\
-e\frac{(1+\tau_{B,3})}{2}A\gamma^0
\right]\psi
=0.
\eeq
Here, we only consider the finite nuclei without the strangeness degree of freedom and the corresponding equations of motion for mesons are given by
\beq
&&-\Delta\sigma+m_\sigma^2\sigma+g_2\sigma^2+g_3\sigma^3+U'_{\text{cut}}(\sigma)
=- g_{\sigma N}\langle\bar\psi_N\psi_N\rangle,\nn
&&-\Delta\omega+m_\omega^2\omega+c_3 \omega^3=g_{\omega N}\langle\bar\psi_N\gamma^0\psi_N\rangle,\nn
&&-\Delta\rho+m_\rho^2\rho=
g_{\rho N}\langle\bar\psi_N\tau_{N,3}\gamma^0\psi_N\rangle,\nn
&&-\Delta A=e\langle\bar\psi_N\frac{(1+\tau_{N,3})}{2}\gamma^0\psi_N\rangle,
\eeq
where, $\tau_{N,3}$ is the third component of nucleon isospin operator and the derivative of $U_{\text{cut}}(\sigma)$ potential is,
\beq
U'_{\text{cut}}(\sigma)=\frac{\alpha\beta g_\sigma}{M_N}\frac{1}{1+\exp[-\beta(g_\sigma\sigma/M_N-f_{s})]}.
\eeq
These coupling equations are solved self-consistently with numerical methods. The ground-state properties of finite nuclei are calculated by using the meson fields and the wave functions of nucleon. Furthermore, in this work, we mainly discuss the properties of doubly-magic nuclei, therefore, the pair effect of nuclei was not considered.

In the infinite nuclear matter system or the core region of neutron star, the nuclear many-body system has the translational invariance. The gradient terms in the RMF Lagrangian will lose their functions.  The Coulomb force also does not play any role due to its divergence in an infinite system. Now the equations of motion of baryons and mesons become as,
\beq
&&\left[\vec\alpha\cdot\vec k+\beta M^*_B
+g_{\omega B}\omega
+g_{\rho B}\rho\tau_{B,3}\gamma^0\
\right]\psi_{Bk}
=\varepsilon_{Bk}\psi_{Bk}
\eeq
and
\beq
&&m_\sigma^2\sigma+g_2\sigma^2+g_3\sigma^3+U'_{\text{cut}}(\sigma)
=-\sum_B g_{\sigma B}
\langle\bar\psi_B\psi_B\rangle,\nn
&&m_\omega^2\omega+c_3 \omega^3=
\sum_Bg_{\omega B}\langle\bar\psi_B\gamma^0\psi_B\rangle,\nn
&&m_\rho^2\rho=
\sum_Bg_{\rho B}\langle\bar\psi_B\tau_{B,3}\gamma^0\psi_B\rangle,
\eeq
where $M^*_B$ is the effective baryon mass related to $\sigma$ field,
\beq
M^*_B=M_B+g_{\sigma B}\sigma
\eeq

From these equations of motion of baryons and mesons, the energy density and pressure are generated by the energy-momentum tensor~\cite{shen02},
\beq
\mathcal{E}&=&\sum_{B}\frac{1}{\pi^2}\int^{k^B_F}_0\sqrt{k^2+M^{*2}_B}k^2dk\nn
&&+\frac{1}{2}m^2_\sigma\sigma^2+\frac{1}{3}g_2\sigma^3+\frac{1}{4}g_3\sigma^4+U'_{\text{cut}}(\sigma)\nn
&&+\frac{1}{2}m^2_\omega\omega^2+\frac{3}{4}c_3\omega^4+\frac{1}{2}m^2_\rho\rho^2
\eeq
and
\beq
P&=&\frac{1}{3\pi^2}\sum_{B}\int^{k^B_F}_0\frac{k^4}{\sqrt{k^2+M^{*2}_B}}dk\nn
&&-\frac{1}{2}m^2_\sigma\sigma^2-\frac{1}{3}g_2\sigma^3-\frac{1}{4}g_3\sigma^4-U'_{\text{cut}}(\sigma)\nn
&&+\frac{1}{2}m^2_\omega\omega^2+\frac{1}{4}c_3\omega^4+\frac{1}{2}m^2_\rho\rho^2.
\eeq

In neutron star matter, there are not only  baryons ($n,~p,~\Lambda,~\Sigma^-,~\Sigma^0,~\Sigma^+,~\Xi^-,~\Xi^0$) but also leptons ($e,~\mu$). All of these particles meet the requirements of charge neutrality and $\beta$ equilibrium. Their chemical potentials should satisfy the following identities,
\beq\label{mueq}
\mu_n&=&\mu_\Lambda=\mu_{\Sigma^0}=\mu_{\Xi^0},\nn
\mu_p&=&\mu_{\Sigma^+}=\mu_n-\mu_e,\nn
\mu_{\Sigma^-}&=&\mu_{\Xi^-}=\mu_n+\mu_e,\nn
\mu_\mu&=&\mu_e,
\eeq
where $\mu_i$ is the chemical potential of particle $i$. They are expressed for baryons $B$ and leptons $l$, respectively
\beq\label{cmq}
\mu_B&=&\sqrt{k^{B2}_F+M^{*2}_B}+g_{\omega B}\omega+g_{\rho B}\tau_{B,3}\rho,\nn
\mu_l&=&\sqrt{k^{l2}_F+m^{2}_l}.
\eeq
Furthermore, the densities of different baryons are limited by the charge neutrality condition,
\beq\label{ceq}
\rho_p+\rho_{\Sigma^+}=\rho_{e}+\rho_{\mu}+\rho_{\Sigma^-}+\rho_{\Xi^-},
\eeq 
where $\rho_i$ is the baryon number density and obtained by 
\beq
\rho_i=\frac{k^{i3}_F}{3\pi^2}.
\eeq 

The properties of a neutron star are obtained from the well-known equilibrium equations by Tolman, Oppenheimer and Volkoff~\cite{tolman39,oppenheimer39} with the pressure $P$ of neutron star matter and the enclosed mass $M$,
\beq\label{tov}
\frac{dP(r)}{dr}&=&-\frac{GM(r)\varepsilon(r)}{r^{2}}\frac{\Big[1+\frac{P(r)}{\varepsilon(r)}\Big]\Big[1+\frac{4\pi r^{3}P(r)}{M(r)}\Big]}
{1-\frac{2GM(r)}{r}},\nn
\frac{dM(r)}{dr}&=&4\pi r^{2}\varepsilon(r),
\eeq
where, $P(r)$ is the pressure of neutron star at radius, $r$, and $M(r)$ is the total star mass inside a sphere of radius $r$. When the EOS $P(\varepsilon)$ is decided from the nuclear many-body method as a function of energy density $\varepsilon$, the total energy density ($G$ is the gravitational constant), the numerical solution of Eq.(\ref{tov}) provides the mass-radius relation of neutron star.

\section{Results and discussions}
Firstly, the TM1 parameter set is chosen as the nucleon-nucleon ($NN$) interaction in RMF model~\cite{sugahara94}, which has achieved a lot of successes in the description of  the structure of nuclear many-body system and the objects of astrophysics. The nonlinear term of $\omega$ meson was firstly introduced in TM1. It generates that the behaviors of scalar and vector potentials in RMF model are consistent with those from an {\it ab initio} method, RBHF theory at high density.  The maximum mass of neutron star is around $2.2M_\odot$ without hyperon within TM1 interaction.  When the hyperons are concerned, the maximum mass of neutron star reduces to $1.6M_\odot$, which is much less than the constraint of massive neutron stars observed recently~\cite{shen02,demorest10,fonseca16,antoniadis13}.  The consideration of $\sigma$-cut potential contributes to  make the EOS stiffer to increase the maximum mass of neutron star. The parameters $\alpha=m^4_\pi$ and $\beta=120$ in $\sigma$-cut potential are taken the same values with those in Ref.~\cite{maslov15}, where the factor $f_s$ was taken as $0.36,~0.44$ and $0.52$.  In the present work, the factor $f_s$ is treated as a free parameter determined by the properties of finite nuclei and neutron star.

In Table \ref{tab1}, the total energies and charge radii of  \ce{^{16}O},  \ce{^{90}Zr}, and  \ce{^{208}Pb}  are listed in terms of different choices of $f_s$ from $0.50$ to $0.60$ and are compared with the original results obtained within TM1 parameter set. It is found that when the factor $f_s$ is larger than $0.55$, the results with $\sigma$-cut potential are identical with those from original TM1 interaction. The smaller $f_s$ corresponds that the $\sigma$-cut potential plays its effect from lower density. Actually, the properties of finite nuclei in RMF model are usually determined by the magnitude of mean-field potential with the densities less than $\rho=0.20$ fm$^{-3}$. The discrepancy between the results with $f_s=0.50$ and the TM1 shows that the $\sigma$-cut potential with $f_s=0.50$ takes its effect below that density. Therefore, to keep the properties of finite nuclei in the present framework without the influence of the $\sigma$-cut potential, $f_s$ must be larger than $0.55$. In the later discussion, we only take $f_s$ as $0.55$ and $0.60$.  

\begin{table}[H]
	\centering
	\begin{tabular}{l c c c c c}
		\hline
		\hline
		Nuclei             &                              &     TM1             ~~~      &  $f_s=0.50$ ~~~       &    $f_s=0.55$      ~~~              &    $f_s=0.60$      \\
		\hline
   \ce{^{16}O} &    $E$    (MeV)  ~~~    &  $-130.3678$ ~~~    &  $-130.2902$~~~      &   $-130.3678$ ~~~             &   $-130.3678$    \\		
		                  &    $r_c$ (fm)     ~~~    &  $2.6589$ ~~~          &        $2.6593$~~~       &   $2.6589$~~~                    &   $2.6589$           \\	
	\hline	                  	
  \ce{^{90}Zr} &    $E$    (MeV)  ~~~   &  $-783.5024$ ~~~   &  $-783.3484$~~~      &   $-783.5024$ ~~~             &   $-783.5024$    \\		
                          &    $r_c$ (fm)     ~~~    &  $4.2634$ ~~~          &        $4.2637$~~~       &   $4.2634$~~~                    &   $4.2634$    \\	
     \hline                    	
  \ce{^{208}Pb} &    $E$    (MeV)  ~~~   &  $-1637.8920$ ~~~ &  $-1637.8487$~~~      &   $-1637.8920$ ~~~             &   $-1637.8920$    \\		
                             &    $r_c$ (fm)     ~~~    &  $5.5311$ ~~~          &        $5.5311$~~~       &   $5.5311$~~~                    &   $5.5311$    \\		
   		  \hline
		\hline
	\end{tabular}
	\caption{The total energies and charge radii of \ce{^{16}O}, \ce{^{90}Zr},  and \ce{^{208}Pb} obtained with TM1 parameter set and within the addition of $\sigma$-cut potentials in term of different $f_s$ factors.}\label{tab1}
\end{table}

\begin{figure}[H] 
	\centering
	\includegraphics[width=11cm]{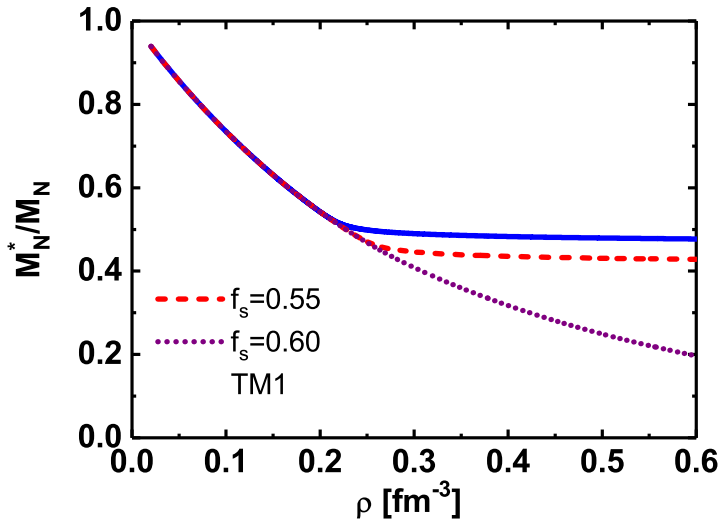}
	\caption{The effective nucleon masses in symmetric nuclear matter as functions of density with the original TM1 and the $\sigma$-cut potentials with $f_s=0.55$ and $0.60$.}
	\label{emr}
\end{figure}

The $\sigma$-cut potential influences the properties of nuclear many-body system through the scalar meson. In RMF model, the effective nucleon masses are defined as $M^*_N=M_N+g_\sigma\sigma$. In Fig.~\ref{emr}, the effective nucleon masses in symmetric nuclear matter as functions of nucleon density obtained with $\sigma$-cut potential and original TM1 parameter set are plotted. The $M^*_N$ decreases monotonously with density increasing in the original TM1 parameter set, which is similar as the effective masses in other interactions of RMF model. Once the $\sigma$-cut potentials are included, the effective masses almost become constants above the certain densities around $0.23-0.27$ fm$^{-3}$, depending on different values of $f_s$. The smaller $f_s$ takes effects earlier and generates the larger effective nucleon masses at high density.  It denotes that the nuclear media effects will be saturated when the nucleon system is highly compact, since the proton and neutron have finite sizes. In other nuclear many-body methods, for example, BHF theory~\cite{baldo14}, quark meson-coupling model~\cite{guichon08} and quark mean field model~\cite{xing16}, the effective nucleon masses also displayed such saturated behaviors in high density region.

\begin{figure}[H]
	\centering
	\subfigure[]{\includegraphics[width=8cm]{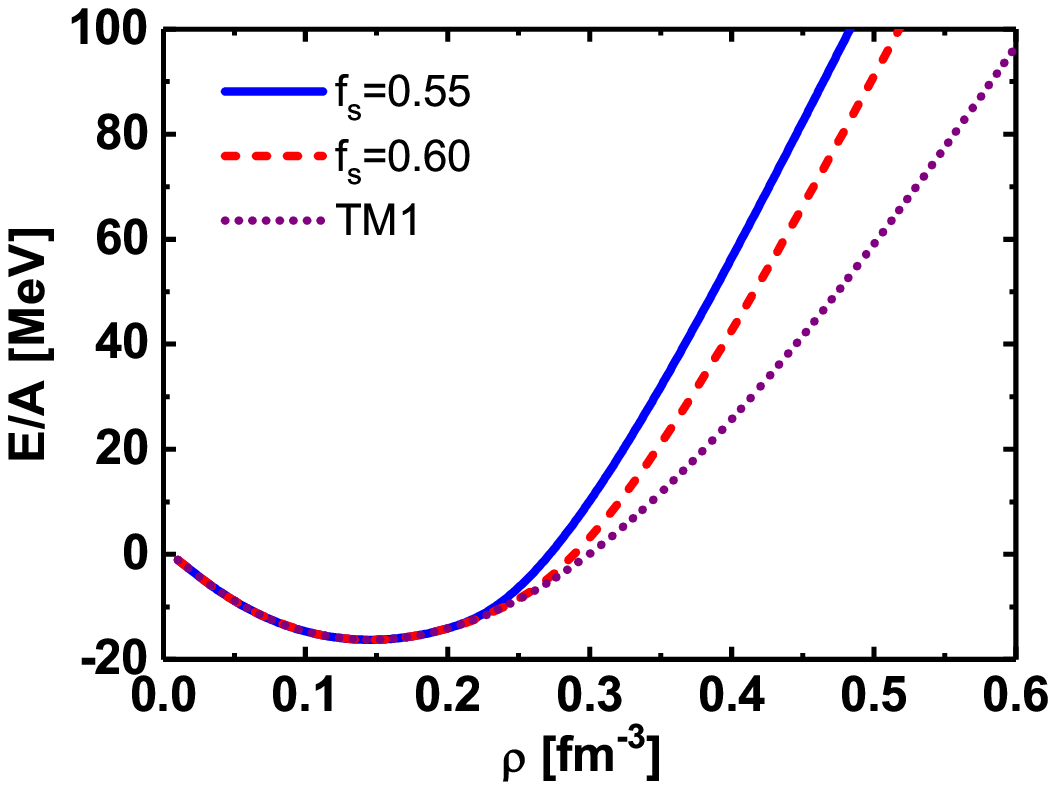}}
	\subfigure[]{\includegraphics[width=8cm]{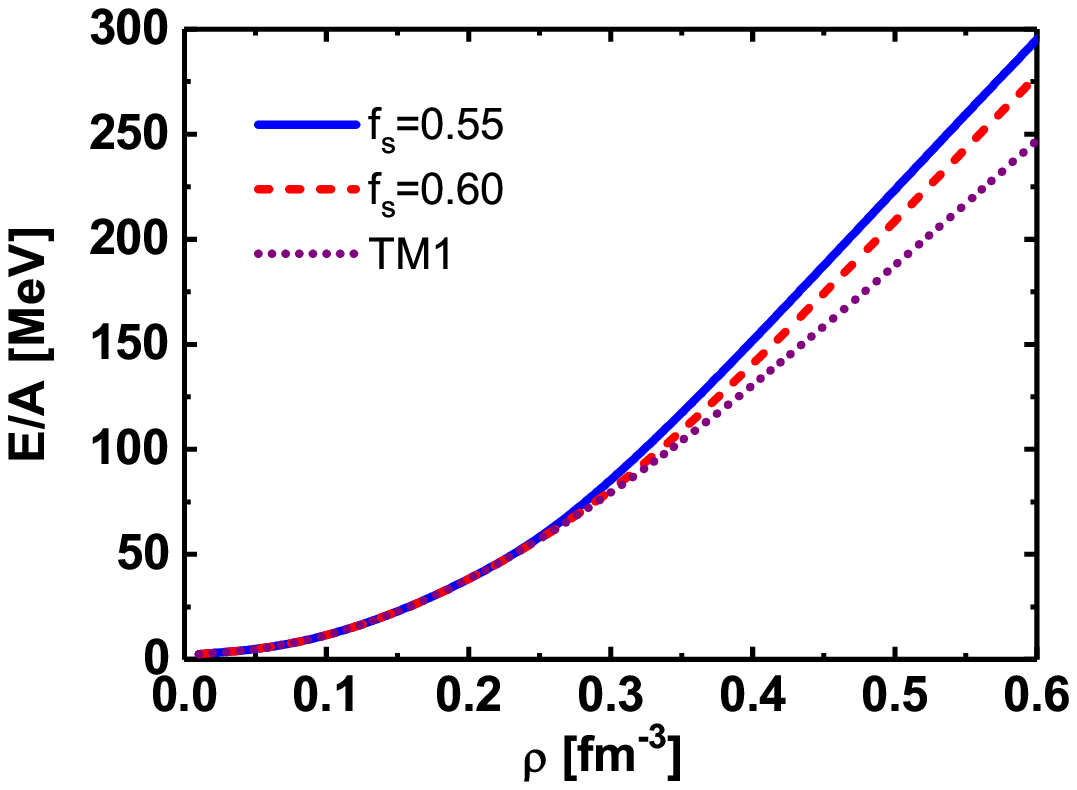}}
	\caption{The binding energies per nucleon as functions of density for symmetric nuclear matter (panel (a)) and for pure neutron matter (panel (b)) with the original TM1 and the $\sigma$-cut potentials with $f_s=0.55$ and $0.60$.}
	\label{eos}
\end{figure}

In Fig.~\ref{eos}, the EOSs of symmetric nuclear matter, panel (a), and pure neutron matter, panel (b), are shown with different $f_s$ in the $\sigma$-cut potential and are compared to those from TM1. The EOSs obtained by considering the high-density cut-off are stiffer than the results from TM1 both in symmetric nuclear matter from $\rho_N=0.23$ fm$^{-3}$ and pure neutron matter from $\rho_N=0.27$ fm$^{-3}$ . The magnitude of $\sigma$ field is reduced by the $\sigma$-cut potential, while the one of $\omega$ field is not changed, which brings the more repulsive contributions to the EOSs and makes them harder at high density. In symmetric nuclear matter, the effect of $\sigma$-cut potential is more obvious than that in pure neutron matter. For example, at $\rho_N=0.40$ fm$^{-3}$, the binding energy from $\sigma$-cut potentials is larger about $80$ MeV than that in TM1, while this amplitude is about $30$ MeV in pure neutron matter. At high density, the $\sigma$ meson fields did not reduce anymore and became saturated in present framework. The $\sigma$-cut potential is isospin independent now, which generates the same strength of $\sigma$ fields in symmetric nuclear matter and pure neutron matter at high density. On the other hand, the $\sigma$ fields in pure neutron matter are smaller than those in symmetric nuclear matter at a certain density without $\sigma$-cut potential. Therefore, the effect of $\sigma$-cut potential on pure neutron matter is weaker than that on symmetric nuclear matter. 

\begin{figure}[H]
	\centering
	\includegraphics[width=11cm]{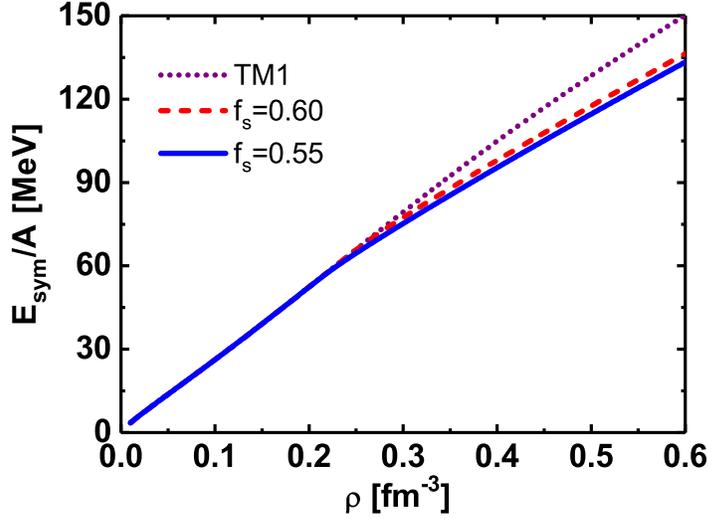}
	\caption{The symmetry energies per nucleon as functions of density with the original TM1 and the $\sigma$-cut potentials with $f_s=0.55$ and $0.60$.}
	\label{sym}
\end{figure}

The symmetry energy is one of the most essential features of nuclear physics, which represents the variation of the binding energy with isospin ~\cite{libaoan08,baldo16}. In Fig.~\ref{sym}, the symmetry energies are shown as functions of density in present framework. Those provided by the $\sigma$-cut potentials are smaller than that generated by TM1 above $\rho=0.25$ fm$^{-3}$. Furthermore, the stronger cut-off corresponding to $f_s=0.55$, provides the smaller symmetry energy. It is caused by that a larger effective mass with $f_s=0.55$ generates a smaller  symmetry energy, since in RMF model, one has the relation about symmetry energy,
\beq
E_{sym}/A=\frac{k^2_F}{6\sqrt{k^2_F+M^{*2}}}+\frac{g^2_\rho}{8m^2_\rho}\rho.
\eeq 

\begin{figure}[H]
	\centering
	\includegraphics[width=11cm]{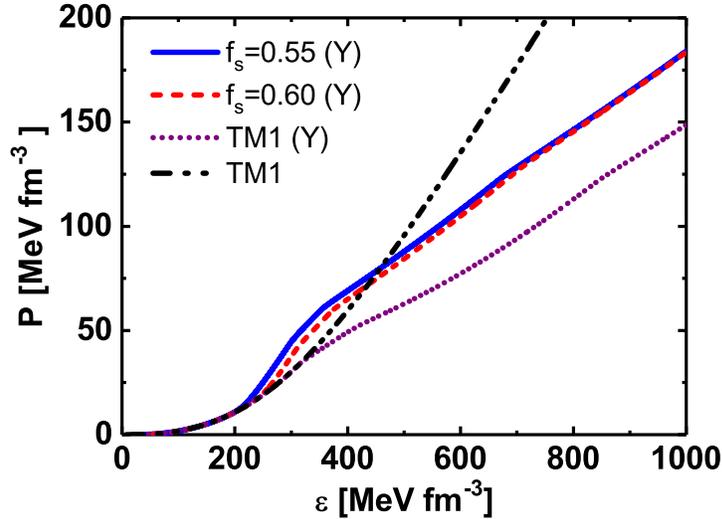}
	\caption{The pressures as functions of energy density with different $\sigma$-cut potentials in neutron star including hyperons and without hyperon and $\sigma$-cut potential. ``Y" in parenthesis means the results including hyperons.}
	\label{ep}
\end{figure}

To discuss the role of strangeness degree of freedom in neutron star, the hyperons, like $\Lambda,~\Sigma$, and $\Xi$, are included in this work. The coupling constants between $\Lambda$ hyperon and mesons are usually fixed by the experimental observation of $\Lambda$ hypernuclei. Due to the lack of experimental information about $\Sigma$ and $\Xi$ hypernuclei, the coupling constants between $\Sigma$ and $\Xi$ hyperons and mesons still have a great deal of ambiguities. Recently, Fortin {\it et al.} systematically studied the neutron star maximum masses constrained by the existing hypernuclei properties in RMF model and discussed the coupling constants between hyperons and mesons in detail~\cite{fortin17}.   In this work, we will follow their choices. The coupling constants between vector mesons and hyperons are given by SU(6) symmetry and those between scalar mesons and hyperons are generated by the empirical hyperon-nucleon potentials, $U^{(N)}_\Lambda=-30$ MeV,  $U^{(N)}_\Sigma=0$ MeV, $U^{(N)}_\Xi=-14$ MeV at
nuclear saturation density, $\rho_0$, and the $\Lambda$-$\Lambda$ potential, $U^{(\Lambda)}_\Lambda=-5.9$ MeV at $\rho_0/5$ in pure $\Lambda$ matter.
\beq
&&g_{\sigma\Lambda}=0.621g_{\sigma N},~~g_{\sigma\Sigma}=0.534g_{\sigma N},~g_{\sigma\Xi}=0.308g_{\sigma N},\nn
&&g_{\omega N}=\frac{3}{2}g_{\omega \Lambda}=\frac{3}{2}g_{\omega \Sigma}=3g_{\omega \Xi},\nn
&&g_{\rho N}=\frac{1}{2}g_{\rho \Sigma}=g_{\rho \Xi},~~~g_{\rho \Lambda}=0,\nn
&&g_{\sigma^*\Lambda}=0.557g_{\sigma N},~g_{\sigma^* \Sigma}=g_{\sigma^* \Xi}=g_{\sigma^* N}=0,\nn
&&g_{\phi \Lambda}=\frac{\sqrt{2}}{3}g_{\omega N},~~g_{\phi \Sigma}=g_{\phi \Xi}=g_{\phi N}=0,
\eeq
where, the strangeness mesons, $\sigma^*$ and $\phi$, are only considered to be exchanged between $\Lambda$ hyperons.  After solving the Eqs. (\ref{mueq}) and (\ref{ceq}) about the $\beta$ equilibrium and charge neutrality conditions of whole system, the relations between pressures and energy densities are given in Fig.~\ref{ep}.  Due to the introduction of $\sigma$-cut potential, the pressures largely increase at lager energy densities comparing with that from the original TM1 interaction, which represents the EOS at high density becomes stiffer leading to a larger mass of neutron star.  Furthermore, the EOS without hyperons and $\sigma$-cut potential is also shown to be compared. Its behavior is similar with those including the $\sigma$-cut effect below the energy density, $\varepsilon=400$ MeVfm$^{-3}$ and becomes harder at high energy density.

\begin{figure}[H]
	\centering
	\subfigure[]{\includegraphics[width=8cm]{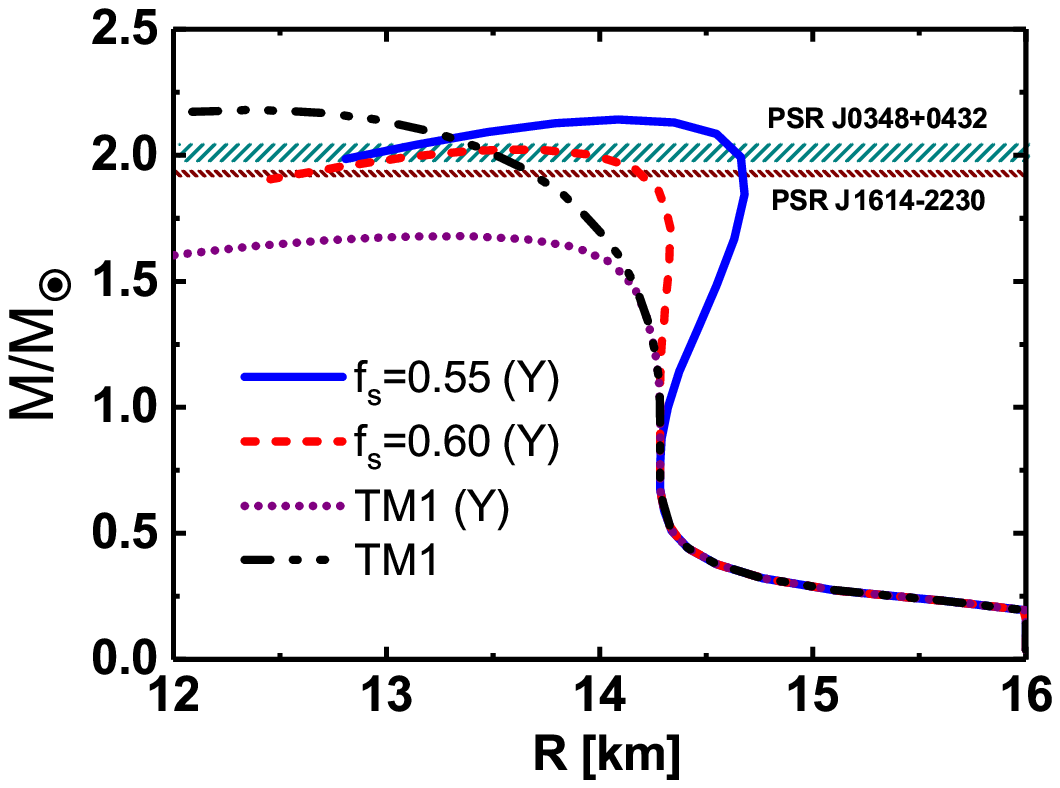}}
	\subfigure[]{\includegraphics[width=8cm]{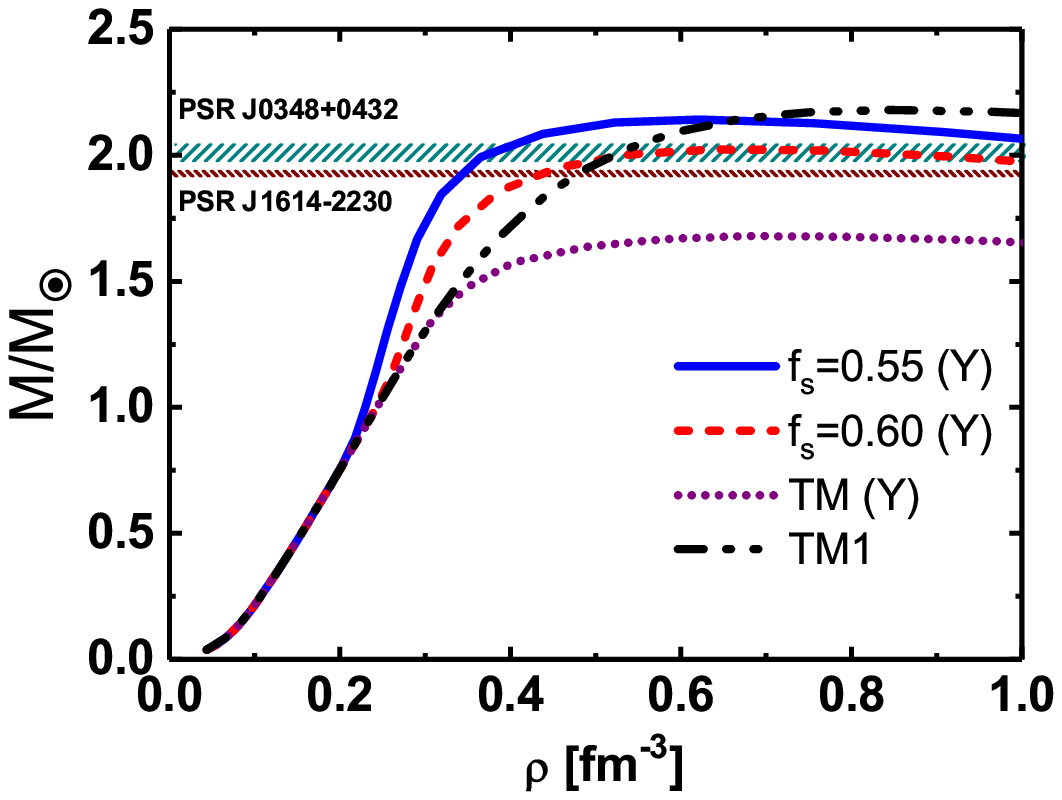}}
	\caption{(a)the mass-radius relation and (b) mass-density relation for neutron star with the original TM1 (with and without hyperons) and the $\sigma$-cut potentials at $f_s=0.55$ and $0.60$. ``Y" in parenthesis means the results including hyperons.}
	\label{rm}
\end{figure}

With the pressures as functions of energy density, the properties of neutron star can be obtained by solving the TOV equation, Eq.~(\ref{tov}). In Fig.~\ref{rm}, the mass-radius and mass-density relations are given in the  panel (a) and panel (b), respectively. The maximum mass of neutron star in TM1 parameter set with hyperons is around $1.68M_\odot$. Once the $\sigma$-cut potentials are included, the maximum masses of neutron star significantly grow up to above $2.0M_\odot$. For the stronger $\sigma$-cut potential $f_s=0.55$, the maximum mass of neutron star approaches $2.14 M_\odot$ and the corresponding radii is about $14.1$ km. However, if $f_s$ is larger than $0.6$, the maximum mass of neutron star will be less than $2.0M_\odot$ which cannot describe the observations of two massive neutron stars, PSR J1614-2230 and PSR J0348+0432. Together with the constraints by the $2M_\odot$ neutron stars without changing the properties of finite nuclei, it can be concluded that the factor $f_s$ in the $\sigma$-cut potential should lie between $0.55$ and $0.60$. Otherwise, the theoretical results could not satisfy the experimental data about the nuclear many-body system. Furthermore, the center densities of neutron stars with $\sigma$-cut potentials are around $0.62$ fm$^{-3}$, which are smaller than that of TM1, due to the stiffer EOSs. 
In Table~\ref{tab2}, the maximum masses, corresponding radii and central densities of neutron stars are tabulated. If the $\sigma$-cut potential was not considered, the maximum mass of neutron star with hyperons is largely reduced from $2.18M_\odot$ without hyperons to $1.68M_\odot$, while they increase to above $2M_\odot$,  when the $\sigma$-cut potentials are included. The corresponding radii become larger and the central densities become smaller.  Furthermore, the radii of neutron stars at $1.4M_\odot,~R_{1.4},$  are also given in this table, which are located around $14.0-14.5$ km.  These values approach the recent constraint by Lattimer and Prakash~\cite{lattimer16}, $9~\text{km}<R_{1.4}<14~\text{km}$. The $R_{1.4}$ in the $\sigma$-cut potential with $f_s=0.6$ are very similar with that in TM1 without hyperons and larger than the one in TM1 with hyperons. This is because that the pressure in the $\sigma$-cut potential with $f_s=0.6$ is very similar with that in TM1 without hyperons in the low energy density region, but larger than the one in TM1 with hyperons as shown in Fig.~\ref{ep}, and the $R_{1.4}$ has a strong correlation with the pressure of neutron star matter at saturation density as pointed in Ref.~\cite{lattimer16}. The inclusion of $\sigma$-cut potential should generate a larger $R_{1.4}$, due to the stiffer EOS as shown in Fig.~\ref{ep}. 
\begin{table}[H]
	\centering
	\begin{tabular}{l  c c c c}
		\hline
		\hline
		                                                          &     TM1 ~~~      &    TM1 (Y)~~~     &    $f_s=0.55$(Y)      ~~~    &    $f_s=0.60$(Y)      \\
		\hline
		$M_\text{max}$  ($M_\odot$)    &  $2.18$ ~~~    &  $1.68$~~~        &   $2.14$ ~~~              &   $2.02$    \\		
		\hline	                  			
		$R_\text{max}$  (km)                   &  $12.37$ ~~~   &  $13.37$~~~      &   $14.09$ ~~~           &   $13.70$    \\			
		\hline                    	
		$\rho_\text{max}$ (fm$^{-3}$) &  $0.85$ ~~~     &  $0.68$~~~        &   $0.62$ ~~~             &   $0.63$    \\			
		\hline                    	
		$R_{1.4}$  (km)                             &  $14.20$ ~~~   &  $14.20$~~~       &   $14.50$ ~~~             &   $14.30$    \\			
		\hline
		\hline
	\end{tabular}
	\caption{The various properties (maximum masses, corresponding radii, and central densities) of the neutron stars with original TM1 parameter set (without and with hyperons)  and the $\sigma$-cut potentials with $f_s=0.55$ and $0.60$. $R_\text{1.4}$ represents the radius of neutron stars at $M=1.4 M_\odot$}\label{tab2}
\end{table}

\begin{figure}[H]
	\centering
	\includegraphics[width=8cm]{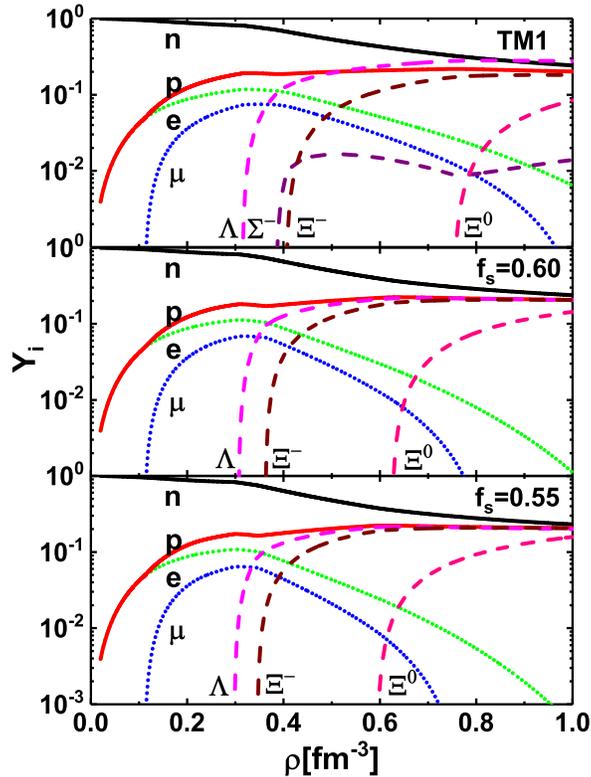}
	\caption{The particle fractions in neutron star with the original TM1 and the $\sigma$-cut potentials with $f_s=0.55$ and $0.60$. }
	\label{yn}
\end{figure}

Finally, the particle fractions in neutron star with different high density cut-offs are displayed in Fig.~\ref{yn}. In TM1, $\Lambda$ hyperon appears first in the core region of neutron star at $\rho_B=0.32$ fm$^{-3}$, which has the deepest hyperon-nucleon potentials among $\Lambda,~\Sigma$ and $\Xi$ hyperons at nuclear saturation density. The appearances of hyperon are determined by their chemical potentials at $\beta$ equilibrium. In RMF theory, the chemical potentials of baryons are written as Eq.~(\ref{cmq}). Furthermore, the free mass of $\Lambda$ hyperon is also smallest in these three hyperons.  The $\Sigma$ hyperon is a little bit heavier than $\Lambda$ hyperons, which appears after $\Lambda$ hyperon. The other hyperons $\Xi^-$ and $\Xi^0$ appear one by one at the higher densities. 

When the $\sigma$-cut potential is taken into account, the $\Xi^-$ and  $\Xi^0$ hyperons appear earlier than $\Sigma^-$ hyperon, whose appearance density is larger than $1.0$ fm$^{-3}$. The onset density of hyperon is above the density where the $\sigma$ terms are turned on. The $\sigma$ dependent potential affects the order of appearance of the hyperons through effective masses of baryons in the chemical potential (Eq.~(\ref{cmq})). The $\sigma$-cut potential leads to larger effective masses of baryons and thus larger corresponding chemical potential. The appearance of hyperons should be retarded. However, the repulsive contribution from the coupling between $\omega$ meson and $\Xi$ hyperon is smaller than that from $\Sigma$ hyperon, which  leads to a much less repulsive potential at high densities. Therefore, the $\Xi$ hyperons will arise in advance. It is also in accordance with the attractive potential between $\Xi$ hyperons and nucleons at nuclear saturation density.

\section{Conclusions}
A $\sigma$-cut interaction was included in the Lagrangian of RMF model with a logarithmic form as a function of $\sigma$ meson field, which can largely reduce the attractive contributions of $\sigma$ meson at high density but does not play any role at low density. There were three parameters in this $\sigma$-cut potential, $\alpha, ~\beta$ and $f_s$. In this work, we focused on discussing the strengths of the factor $f_s$ without changing the ground-state properties of finite nuclei with original RMF interaction and the constraints of observed massive neutron stars.

The binding energies and charge radii of  \ce{^{16}O},  \ce{^{90}Zr}, and  \ce{^{208}Pb} were calculated within the TM1 parameter set and the $\sigma$-cut potentials with different $f_s$. It was found that the $f_s$ should be larger than $0.55$ so that the additional $\sigma$-cut potentials in RMF model cannot influence the accurate description of the finite nuclei system. Furthermore, the properties of symmetric nuclear matter and pure neutron matter were also investigated, such as effective nucleon mass, binding energy per particle and symmetry energy. The smaller $f_s$ made the $\sigma$-cut potential take its effects earlier and led to a stronger repulsion. The effective nucleon masses were saturated at high density, which was regarded as the saturation character of nucleon media effect at highly compact system and was consistent with the conclusions from the Brueckner-Hartree-Fock method and quark meson-coupling model. The EOSs of symmetric nuclear matter and pure neutron matter became stiffer at high density due to the $\sigma$ field reduction.  The symmetry energies within $\sigma$-cut potentials were smaller than that from TM1, since they are determined by the effective nucleon masses in RMF model. The larger effective mass provides smaller symmetry energy.

In the last part, the properties of neutron star were studied within the present framework including the strangeness degree of freedom. The maximum masses of neutron star increased from $1.68 M_\odot$ to above $2.0 M_\odot$ when the $\sigma$-cut potential were used with the factor $f_s$ smaller than $0.60$. In this way, the hyperons may exist in the core region of massive neutron stars whose masses are around $2M_\odot$. The $\Xi$ hyperons appeared earlier with the $\sigma$-cut potential comparing to the original TM1 interaction.   

Therefore, with the constraints of finite nuclei and massive neutron stars, the strengths $f_s$ in the $\sigma$-cut potential should be between $0.55$ and $0.60$.  Through including a simple logarithmic interaction, the properties of finite nuclei and massive neutron stars with hyperons can be both described reasonably.  However, the properties of nuclear matter at high density need to be further investigated with more fundamental nuclear many-body theories due to the lack of experimental constraints.

\section{Acknowledgments}
J. Hu would like to thank Dr. Hong Shen for the useful discussion on the calculations of neutron star within RMF model. This work was supported in part by the National Natural Science Foundation of China (Grants No. 11405090, No. 11405116 and No. 11775119) and the Fundamental Research Funds for the Central Universities.

\end{document}